\documentclass{aa}  

\usepackage{graphicx}
\usepackage{txfonts}
\usepackage{lipsum}
\usepackage{subcaption}         
\usepackage{lscape}             
\usepackage{placeins}           
\usepackage{amssymb} 
\usepackage{amsmath} 
\usepackage{color}

\def\kms{{\text{km\,s}$^{-1}$}\xspace}

\def\Lsun{{\rm L$_{\odot}$}}
\def\Rsun{{\rm R$_{\odot}$}}
\def\Msun{{\rm M$_{\odot}$}}

\def\he2{{He~{\small II}}\xspace}
\def\ebv{ {$E({\rm B}-{\rm V})$}\xspace }

\begin{document}

   \title{Dimming and pulsation shock of the coalesced star V838 Monocerotis}

   \author{T. Kamiński\inst{\ref{inst1}}
   \and C.~E. Woodward\inst{\ref{inst2}} 
   \and T. Liimets\inst{\ref{inst3}}
   \and M.~R. Schmidt\inst{\ref{inst1}}
   \and A. A. Djupvik\inst{\ref{inst4}}\inst{\ref{inst5}} 
   \and I. Ilyin\inst{\ref{inst6}} 
        }

   \institute{
   Nicolaus Copernicus Astronomical Center, Polish Academy of Sciences, Rabia{\'n}ska 8, 87-100 Toru{\'n}, Poland \email{tomkam@ncac.torun.pl} \label{inst1}
   \and Minnesota Institute for Astrophysics, School of
   Physics \& Astronomy, 116 Church Street SE, University of Minnesota,
   Minneapolis, MN 55455, USA  \label{inst2}
   \and Tartu Observatory, University of Tartu, Observatooriumi 1, T\~oravere 61602, Estonia  \label{inst3}  
   \and Nordic Optical Telescope, Rambla Jos\'{e} Ana Fern\'{a}ndez P\'{e}rez 7, ES-38711 Bre\~{n}a Baja, Spain  \label{inst4}
   \and Department of Physics and Astronomy, Aarhus University, Munkegade 120, DK-8000 Aarhus C, Denmark  \label{inst5}
   \and Leibniz-Institute for Astrophysics Potsdam (AIP), An der Sternwarte 16, D-14482 Potsdam, Germany  \label{inst6}
   }


 
  \abstract
   {V838 Monocerotis is the remnant of a stellar merger that occurred in 2002. Twenty-four years after the merger, the remnant closely resembles a red supergiant, but its luminosity is sustained by core hydrogen burning and continued contraction toward hydrostatic equilibrium. In late 2025, the system entered the deepest dimming event observed since 2006.}
   {We characterize the 2026 dimming event of V838 Mon using multiband photometry and high-resolution spectroscopy spanning from the dimming minimum through the recovery phase, and investigate whether the merger remnant exhibits pulsations analogous to those seen in red supergiants and Mira stars.} 
   {We obtained $UBVRI$ and $JHK_s$ photometry with the NOT telescope and extracted light curves from the public AAVSO, ATLAS, ZTF, and ASAS-SN databases and services, covering the entire dimming event. These data were supplemented by time-series spectra acquired with SALT/HRS, LBT/PEPSI, VLT/Xshooter, and VLT/UVES, covering the spectral range 3760--24\,800 \AA\ at resolutions $R=6\,700$--130\,000. Differential spectra were constructed by comparison with pre-dimming spectra from 2024.}  
   {{{The photometric color evolution during the dimming can be well reproduced by grey dust extinction with $A_V$ = 1.3$\pm$0.1 mag, consistent with a transiting clump of freshly formed circumstellar dust.}} The photospheric effective temperature changed by no more than $\approx$200 K during the event. During the recovery phase, hydrogen recombination lines from Balmer, Paschen, and Brackett series appeared in emission.{{Their line ratios do not follow the Case B standard but are reminiscent of hydrogen emission in pulsating M-type Mira stars near maximum light. These features are interpreted as arising from a pulsation shock propagating outward from below the photosphere.}} Simultaneously, low-ionization metal lines appeared blueshifted by 90 \kms\ relative to the stellar rest frame, tracing shock-accelerated gas on the near side of the stellar disk. The spectroscopic sequence suggests that the 2026 dimming was itself triggered by a preceding pulsation shock that occurred earlier in 2025 when the star was in conjunction with the Sun.}  
   {We present the first observational evidence for pulsations in a stellar merger remnant. Twenty-four years after the coalescence, V838 Mon exhibits pulsation shocks closely resembling those seen in red supergiants and Mira stars, providing observational support for predicted pulsational instability in post-merger objects. The observed shock may trigger a subsequent dimming event later in 2026.}


   \maketitle
\nolinenumbers

\section{Introduction}\label{sect-intro}
In 2002 V838 Mon underwent a red nova outburst \citep{Munari2002,TylendaEvolutionV838}, which is thought to have been a merger \citep{TylSoker2006} in a hierarchical triple system \citep{TylendaSpec2009,KamiALMA}. In many respects, the outburst was prototypical of red novae, but V838 Mon remains the only object of this class to have had a very young progenitor, most likely a 6--10 \Msun\  B3 main-sequence star that merged with a 0.3 \Msun\ protostar \citep{KamiBlago}. The remnant of the merger is an $\sim$M3 type star with high luminosity ($2.3\times10^4$ \Lsun), huge radius ($\approx$460 \Rsun), convective envelope \citep{KamiLit}, clumpy, dust- and molecule-rich wind \citep{KamiALMA} with SiO masers \citep{Ortiz-Leon}. These characteristics make it very similar to red supergiants with high mass-loss rates and to some massive AGB stars, but unlike these evolved objects, the merger remnant of V838 Mon is powered by core hydrogen burning and by contraction to a hydrostatic equilibrium. The post-merger evolution of V838 Mon relevant to this study has been reviewed in \cite{TylendaEngulf}, \citet{Loebman},  and \cite{Liimets}. Twenty-four years after the collision, the merger ejecta from the 2002 event is still apparent in emission at millimeter wavelengths and in molecular absorption at shorter wavelengths. Due to the expansion and cooling, this component is becoming visibly weaker in successive observations.

In 2006--2008 the system underwent a dimming phase caused by the merger ejecta reaching and engulfing the surviving B3\,V companion, located at a projected distance of $\approx250$ au from the M star \citep{eclipse2,TylendaEngulf}. After this event, the B3 companion disappeared from the optical spectrum, but its surrounding compact \ion{H}{II} region has remained visible in faint H$\alpha$ and even weaker lines of photoionized metals. The photometric fluxes of the entire system have been steadily increasing since the 2006--2008 event. As will be discussed in a forthcoming paper, this brightening is mainly caused by a reduction in circumstellar extinction toward the system, while the cool stellar remnant remains largely unchanged \citep{Geballe2025}. However, for the last decade or so, the visual bands of the M star have displayed 0.1--0.2 mag variations. They are irregular or quasi-periodic and very similar to the variability displayed by genuine red supergiants, where convection and radial pulsations in the fundamental mode mix with non-radial modes of longer periods \citep{Harper2020}. \cite{goran2020} advocated a period of about 360 d for V838 Mon's variability, close to the 400 d period of Betelgeuse \citep{MontargesDimmingNature}, which, however, has not been confirmed by later Fourier analysis of the V838 Mon light curves \citep{Liimets}. This period is also substantially longer than the dynamical timescale of the coalesced star \citep[$\approx$75 or 58 d, computed from the stellar parameters in][for assumed masses of 6 or 10\,\Msun, respectively]{KamiALMA}, but, admittedly, the photospheric radius is still very uncertain, and the star has not yet reached an equilibrium after the merger. 

\begin{figure}
    \centering
    \includegraphics[width=0.99\columnwidth]{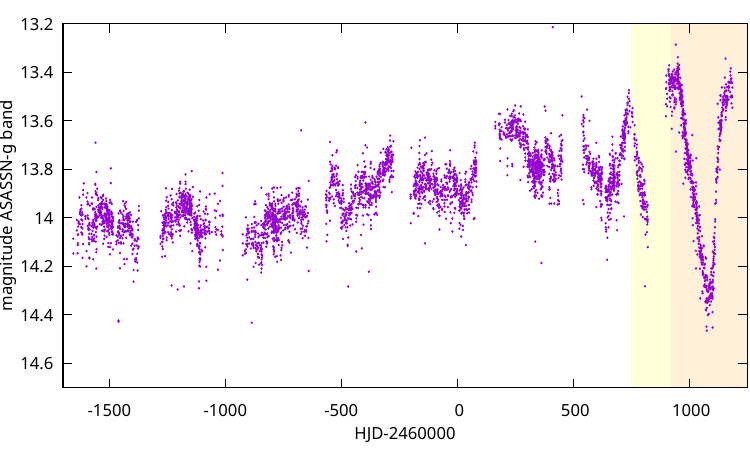}
    \includegraphics[width=0.99\columnwidth]{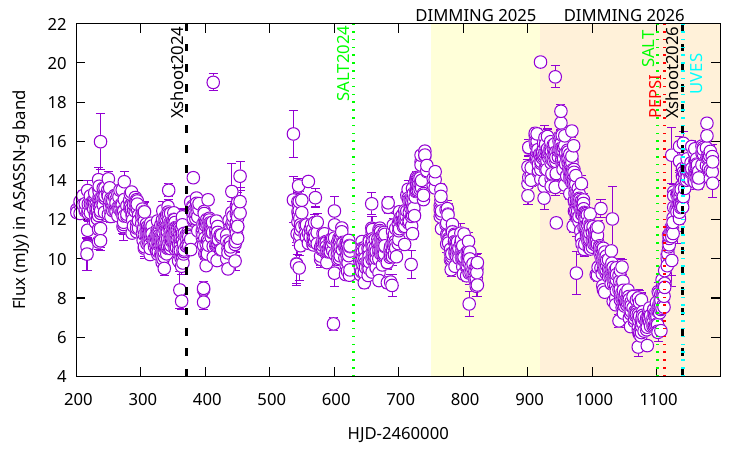}
    \includegraphics[width=0.99\columnwidth]{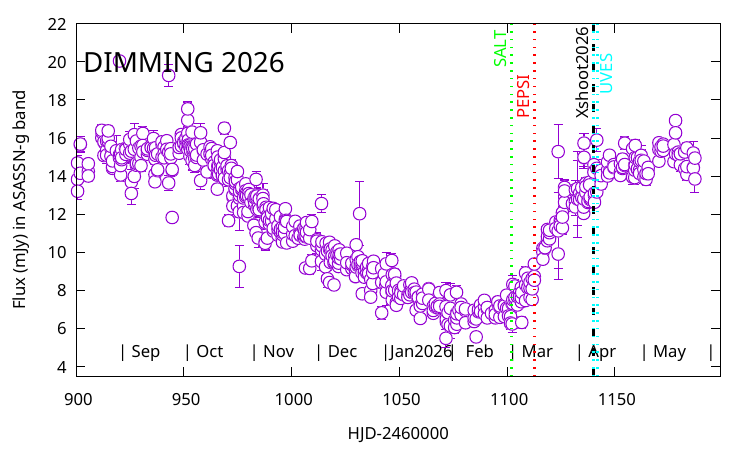}
    \caption{The $g$-band light curve of V838 Mon {{from ASSASN. The top panel shows $g$ magnitudes from Aug 2018 to May 2026. The mid panel shows more recent variability, in flux units. The bottom panel zooms in on the dimming episode with a minimum in Feb. 2026.}} Spectral observations are indicated by vertical lines and labels.}
    \label{fig-lightcurve}
\end{figure}

In October 2025, V838 Mon started a decline in visual bands \citep{Goranskij2026ATel}. Initially, the event resembled earlier brightness declines, particularly the episode observed in early 2025. However, by the end of 2025 it turned out to be the deepest dimming event seen in the system since 2006--2008. The decline in brightness closely resembles dimming events observed in red supergiants \citep{dimmingRWCep}, including the Great Dimming of Betelgeuse \citep{MontargesDimmingNature}. The photometric evolution in the $g$ band from the ASSASN Sky Patrol service \citep{asasCatII} is shown in Fig. \ref{fig-lightcurve}. The light curve reached minimum brightness in February 2026, and the pre-dimming flux level was restored by April 2026. In the $g$ band, the minimum was at $\approx$39\% of the flux levels just before and just after the dimming. Here, we analyze the dimming event using intermediate and high resolution spectra obtained during the recovery phase, i.e., after the flux minimum (Fig. \ref{fig-lightcurve}). While it has been postulated that merger products are prone to pulsation instabilities \citep{RuiFuller,Henneco2025}, it has not been verified observationally for an immediate merger product. Here, we find strong indications that 24 years after the merger, V838 Mon exhibits pulsation shocks, similar to those seen in red supergiants and M-type miras.   

\section{Observations}
The dimming event was covered by photometric and spectroscopic observations. Our photometric measurements are summarized in Table~\ref{tab-not}. The spectral observations are summarized in Table \ref{tab-spectra}. Note the differences in spectral coverages and spectral resolutions ($R=\lambda/\Delta\lambda$). 

\subsection{NOT photometry}\label{sect-obs-phot}
$UBVRI$ photometry on four dates was acquired with the ALhambra Faint Object Spectrograph and Camera (ALFOSC) at the 2.6m Nordic Optical Telescope (NOT). Exposure times for all dates were 500, 150, 10,  2, and 0.3 s for $U$, $B$, $V$, $R$, and $I$, respectively. Data were bias and flat-field corrected using IRAF\footnote{IRAF was distributed by the National  Optical Astronomy Observatory, which was operated by the Association  of Universities for Research in Astronomy (AURA) under cooperative  agreement with the National Science Foundation.} (\citealt{1986SPIE..627..733T,1993ASPC...52..173T}) and aperture photometry obtained with the GAIA Starlink software\footnote{https://astro.dur.ac.uk/~pdraper/gaia/gaia.html}. Instrumental magnitudes were transformed into standard Johnson-Cousins magnitudes using the fixed color terms provided at the ALFOSC webpage\footnote{https://www.not.iac.es/instruments/alfosc/zpmon/} and comparison stars in the field of view, as described in \cite{Liimets}. {{We use the $R$ and $I$ designations for the Cousins magnitudes throughout this paper.}} All measurements, together with respective seeing values, are shown in Table~\ref{tab-not}.

Observations were also taken in the near-infrared (NIR) bands $JHK_s$ with the NOTCam instrument just after the dimming event, on 28 April 2026. 
The observations were obtained as in \citet{Liimets}, using the HR-camera (0\farcs078/pix) and a 6 mm small cold stop to effectively reduce the 2.56 m NOT aperture to a 1 m telescope to avoid saturation. The seeing was around 0\farcs5, meaning the diffraction limit was reached in the $K$-band. Observations were obtained in a 5-point dither for V838 Mon and a nearby standard star,  HD296273, observed right before. Domeflats lamp ON-OFF obtained with the same setup were used for flat-fielding, and the images were then sky-subtracted, shifted, and combined. Photometric magnitudes were measured with large aperture radii (90 pixel = 7\arcsec) for both the bright standard and the target, to be consistent with previous measurements in \citet{Liimets}. 2MASS catalog magnitudes for the standard star were used to calibrate the V838 Mon magnitudes and yielded $J$ = 6.093$\pm$0.02, $H$ = 5.132$\pm$0.03, and $K_s$ = 4.372$\pm$0.03. Within the uncertainties, these values are essentially the same as values obtained on 3 Nov. 2005 with the same instrument and observing techniques, $J$=6.09$\pm$0.02, $H$=5.08$\pm$0.03, and $K_s$=4.36$\pm$0.03 mag.   

\begin{table*}
\caption{NOT optical photometry during the 2026 dimming.}
\label{tab-not}
\centering
\begin{tabular}{llcccccc}
\hline\hline
Date & HJD & $U \pm \sigma_{U}$ & $B \pm \sigma_{B}$ & $V \pm \sigma_{V}$ & $R_{C} \pm \sigma_{R_{C}}$ & $I_{C} \pm \sigma_{I_{C}}$  & Seeing \\
&2460000+& [mag] & [mag] & [mag] & [mag] & [mag] & [$''$]\\
\hline
12-10-2025 & 0961.684106 & 17.85 $\pm$ 0.07 & 15.24 $\pm$ 0.02 & 12.55 $\pm$ 0.01 & 10.76 $\pm$ 0.02 & 8.70 $\pm$ 0.01 & 0.5 -- 0.8 \\ 
02-11-2025 & 0982.721806 & 18.18 $\pm$ 0.07 & 15.50 $\pm$ 0.02 & 12.80 $\pm$ 0.02 & 10.93 $\pm$ 0.01 & 8.80 $\pm$ 0.01  & 0.7 -- 1.1 \\
01-03-2026 & 1101.376362 & 18.17 $\pm$ 0.07 & 16.42 $\pm$ 0.02 & 13.83 $\pm$ 0.02 & 11.68 $\pm$ 0.02 & 9.29 $\pm$ 0.02  & 0.8 -- 1.3 \\
07-03-2026 & 1107.378649 & 18.22 $\pm$ 0.07 & 16.25 $\pm$ 0.01 & 13.64 $\pm$ 0.02 & 11.56 $\pm$ 0.01 & 9.20 $\pm$ 0.01  & 0.7 -- 1.4 \\
\hline
\end{tabular}
\end{table*}

{{
\subsection{Archival and public photometry}
We also used photometry available from public surveys. Aperture photometry in the Sloan $g$ band was obtained directly from the ASASSN Sky Patrol version 1 \citep{asasCatII}. We additionally used multiband photometry submitted to the American Association of Variable Star Observers (AAVSO) database. The 2026 dimming was covered by $BVRI$ measurements in the Johnson--Cousins system. Only mutually consistent measurements were retained, and several outliers in the $B$ and $R$ bands were discarded. On dates when the AAVSO observations overlap with our NOT measurements, the discrepancies between the two datasets exceed their quoted uncertainties. We attribute these systematic differences to slight differences in the effective photometric systems. Data from the Asteroid Terrestrial-impact Last Alert System (ATLAS) survey \citep{atlas1,atlas2} were retrieved through the ATLAS Forced Photometry Server \citep{Shingles} using its Python application programming interface (API). Measurements obtained on the same night were averaged. The ATLAS {\it orange} data showed significant saturation and were therefore discarded. We use only the {\it cyan} ($c$) measurements, some of which may still be mildly affected by saturation.

We also used photometric products from the Zwicky Transient Facility (ZTF). Only a single photometric measurement, in the ZTF $g$ band, was available for the 2026 dimming in the public Data Release 24 (DR24) at IPAC\footnote{\url{https://irsa.ipac.caltech.edu/cgi-bin/Gator/nph-dd}}. More recent ZTF observations were retrieved through the ZTF Forced Photometry Service \citep{masci}. The returned difference-image PSF photometry was filtered using the quality flags and diagnostic quantities recommended in the ZTF documentation, including image-noise and PSF-fit statistics. Total source fluxes were reconstructed by adding the reference-image source flux to the differential flux, and only measurements with $S/N>3$ were retained. Individual measurements were rescaled to a common photometric zeropoint and combined within each observing night using inverse-variance weighting in flux space. The $r$- and $i$-band forced photometry was unusable because V838 Mon is saturated in all available ZTF images.
}}

\begin{table*}
\caption{Summary of spectral observations.}
\label{tab-spectra}
\centering
\begin{tabular}{cc cc c}
\hline\hline
Date & HJD 2460000+ & Instrument & Spectral ranges (\AA) & Resolution $R$ \\
\hline
02-03-2024 & 0371.58 & VLT/Xshooter [UVB; VIS; NIR] & 2988--5559; 5336--10199; 9939--24786 & 6700; 18\,400; 11\,600\\
15-11-2024 & 0629.51 & SALT/HRS [red; blue] & 3827--5574; 5419--8798 & 66\,700; 73\,700 \\
02-03-2026 & 1102.35 & SALT/HRS [red; blue] & 3827--5574; 5419--8798 & 66\,700; 73\,700\\
13-03-2026 & 1112.67 & LBT/PEPSI & 3819--9064 & 130\,000\\
10-04-2026 & 1140.05 & VLT/Xshooter [UVB; VIS; NIR] & 2988--5559; 5336--10199; 9939--24786 & 6700; 18\,400; 11\,600\\
10-04-2026  &  1140.06 & VLT/UVES [blue437; red860] & 3760--4880; 8670--10410&80\,000; 110\,000\\
12-04-2026  &  1142.05 & VLT/UVES [blue437; red860] & 3760--4880; 8670--10410&80\,000; 110\,000\\
\hline
\end{tabular}
\end{table*}

\subsection{SALT spectroscopy}
Observations with the High Resolution Spectrograph \citep[HRS][]{CrauseHRS} on the South African Large Telescope (SALT) telescope were obtained on 02-03-2026 during the dimming, but for reference we also use observations from 15-11-2024 obtained with the same instrument configuration. The high-resolution mode of the spectrograph was used. HRS is a dual-beam (blue and red arm) fiber-fed échelle spectrograph. Separate fiber was used to register sky fluxes 60\arcsec\ away from the science target.  This resulted in imperfect sky correction, and resultant science spectra have residual telluric emission features. Standard calibration frames were obtained, including ThAr arc spectra. Data were processed by the Midas automatic pipeline \citep{HRSpipeline}. Due to sparse arc line coverage in the last few échelle orders of the red arm, the wavelength calibration is suboptimal, with shifts reaching 0.4 \AA. The spectra were only roughly corrected by instrument response functions and are not calibrated in flux. Since SALT's effective aperture changes over the exposure time, flux calibrations are particularly difficult and were not attempted here. 

\subsection{PEPSI spectroscopy}

Observations of V838~Mon were obtained at the $2\times8.4$m Large Binocular Telescope (LBT)  using the Potsdam échelle Polarimetric and Spectroscopic Instrument (PEPSI) \citep[][]{2015AN....336..324S} on 13-03-2026 under photometric conditions with seeing of $\sim$ 0\farcs4 at an airmass of 1.30. PEPSI was configured with the 200~$\mu$m fiber (with an on-sky aperture diameter of 1\farcs50).
Both apertures of the LBT telescope were used and science data were taken simultaneously with that of an off-source sky-fiber.  V838~Mon was observed in all six cross dispersers 
resulting in data that spanned a spectral range from 3819 to 9064 \AA. The total exposure time was $\sim 20$~min.
The raw data were run through the AIP Postdam Spectroscopic Data Systems pipeline \citep{2000PhDT..........I, 2018A&A...612A..44S}. 
The wavelength solution was derived from calibration observations of a ThAr lamp. The spectrum was corrected for the bias, flat field, and scattered light. Extracted échelle orders were merged. The spectrum was not calibrated in flux units.


\subsection{Xshooter}
The XSHOOTER (hereafter Xshooter) spectrograph \citep{xshooter} on the Very Large Telescope (VLT) was used to obtain spectra in a wide range, 3000--24\,000 \AA, near the end of the 2026 dimming, on 10-04-2026. For comparison, we also use here an Xshooter spectrum from 02-03-2024. Xshooter has three arms: UVB, VIS, and NIR. The observations were obtained in service mode following standard Xshooter procedures and the default calibration plan.\footnote{\url{https://www.eso.org/sci/facilities/paranal/instruments/xshooter/doc/VLT-MAN-ESO-14650-4942_P114v1.pdf}}. The 2024 observations were obtained in `stare' mode, whereas in 2026 we used `nod-on-slit' mode to optimize the NIR observations, with a 5\arcsec\ nod throw and a 1\arcsec\ jitter box. In both epochs, the slit widths were set to 0\farcs8 in the UVB arm, and 0\farcs4 in the VIS and NIR arms. To avoid saturation in the red, exposures in 2024 comprised 4, 4, and 28 exposures of 305, 219, and 50 s in the UVB, VIS, and NIR arms, respectively, while in 2026 they were arranged in four cycles of 1, 2, and 6 exposures of 312, 62, and 60 s. Some narrow spectral ranges were saturated in the 2024 NIR spectrum. Seeing (airmass) was 0\farcs95 (2.8) and 0\farcs96 (2.0) in the 2024 and 2026 observations, respectively. All spectra were processed within the Xshooter pipeline \citep{XshooterPipeline} using standard settings. Additionally, reduced spectra were corrected for telluric absorption features with {\tt molecfit} \citep{molecfit2} and using the science spectrum itself for determining the correction.

The Xshooter spectra were flux-calibrated by the pipeline through comparison with spectrophotometric standards observed in the same configuration. However, synthetic photometry derived from the 2024 and 2026 Xshooter spectra differs from our and literature photometry by up to $\sim1$ mag, indicating substantial uncertainty in the absolute flux calibration. This discrepancy is most likely caused by large slit losses for V838 Mon\footnote{\url{https://support.eso.org/en-GB/kb/articles/data-processing-of-x-shooter-data-faqs}}. Relative flux calibration within each Xshooter arm, relevant for emission lines analyzed here, is expected to be much better.
{{The spectrophotometric standards were observed through a much wider slit than V838 Mon, so slit losses for the standards are negligible. The narrow science slits can introduce a substantial multiplicative uncertainty in the absolute flux scale, whereas the associated wavelength dependence is expected to vary smoothly within each Xshooter arm. Consequently, relative fluxes of spectral features within a given arm, particularly over limited wavelength intervals, are expected to be considerably more reliable than the absolute flux calibration.
Slit losses are expected to be smaller in the NIR arm. }}

\subsection{UVES}
The last set of spectra obtained during the dimming event was acquired with Ultraviolet and Visual Echelle Spectrograph (UVES) at VLT \citep{DekkerUVES}. They were taken on 10-04-2026, on the same night immediately after the Xshooter spectrum, and two days later. Both observing runs used the same instrumental setup with dual-beam observations in the `dichronic 437+860' setup with 0\farcs3 slits in blue and red arms and with the image slicer \#3. This resulted in spectral coverage as in Table \ref{tab-spectra}, but with the last four echelle orders having 20--30 \AA\ gaps between them. On 10-04-2026, four exposures of 1574 s were made with each arm at seeing 0\farcs5--0\farcs9 and an airmass of 1.2--1.6. Two days later, two exposures were obtained at a seeing of 0\farcs5--0\farcs6 and an airmass of 1.3--1.4. All calibration observations were obtained according to the standard calibration plan of the observatory. All spectra were processed within the UVES pipeline using standard settings. Within the noise levels, the UVES spectra from the two different nights were identical and were averaged for better S/N. 

\subsection{Notes on all spectral observations}
Given the very steep slope of the ISM-reddened spectral energy distribution (SED) of V838~Mon, the signal-to-noise ratio (S/N) was always very satisfactory at wavelengths long-ward of about 4500~\AA, but often no flux was registered below the noise level at $\lambda \lesssim$4000 \AA. As in Xshooter spectra, the peak flux density is reached near 1.7 $\mu$m before the spectra are corrected for ISM extinction. To compare the quality of all the extracted spectra, in Table \ref{tab-sn} we list their S/Ns near 4500 and 7000 \AA, where S/N is calculated as the pseudocontinuum flux divided by the noise rms at that wavelength; true noise in the spectra is often difficult to measure due to omnipresent molecular features, especially in the high-S/N spectra of Xshooter, and so the derived S/Ns, especially for the red part, are often underestimated. Values are given in their native binning and resolution. 
For most parts of our analysis, we used spectra corrected for interstellar extinction with \ebv=0.9 mag and $R_V$=3.1 as in \citet[][]{TylendaProgenitor}. All reported velocities in this study are in the heliocentric rest frame. The systemic radial velocity of the M star V838 Mon is 75$\pm$1 \kms, as measured from the SiO maser surrounding the merger remnant \citep{Deguchi2005,Ortiz-Leon}.

\subsection{Differential spectra}
The photospheric spectrum of V838 Mon, like all spectra of M3 giants, is particularly rich in spectral features with no regions of line-free continuum. To better explore dimming-related features in the busy 2026 spectra, we compared the SALT and Xshooter spectra from the dimming event to corresponding spectra obtained in 2024 with the same instrument and slit widths. For Xshooter, the spectra were acquired with different observing schemes, but this has no practical consequences for the discussion here. 
After sigma-clipping and scaling the spectra by their respective median values every 100 \AA, we obtained a differential spectrum, 2026 minus 2024, which makes changes particularly apparent. The difference spectrum is
$$ S_{\rm diff}=S_{2026}-S_{2024}\times\frac{median_{\rm 100A}(2026)}{median_{\rm 100A}(2024)}\approx S_{\rm 2026}-2\times S_{2024}.$$
There were no suitable archival UVES and PEPSI data for such a differential analysis from before the dimming, but our spectra from the dimming were compared to each other in a similar fashion as described above (with necessary smoothing to the same resolution).

\begin{figure*}
\sidecaption
\includegraphics[width=11cm]{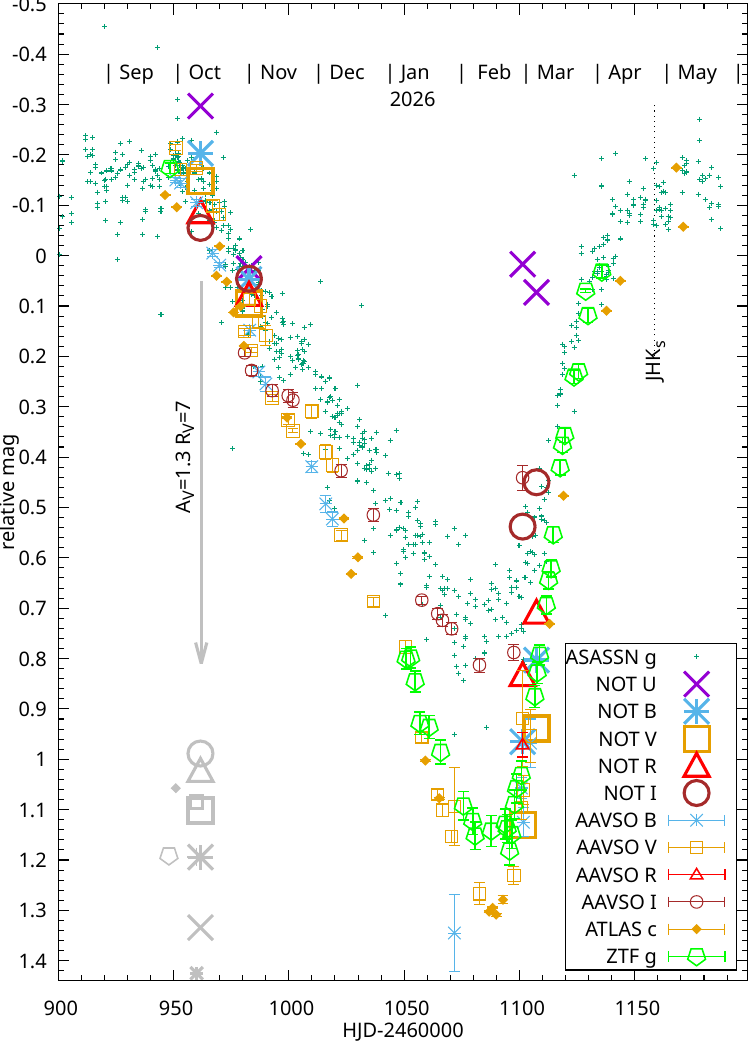}
\caption{{Color evolution during the 2026 dimming of V838 Mon. Our NOT data (large symbols) and public AAVSO, ATLAS, ZTF, and ASASSN data are included. Photometric points were arbitrarily normalized for better clarity (NOT data were normalized to the second epoch of NOT data, near day 983). 
The subtracted $UBVRI$ magnitudes from NOT data are 18.15, 15,45, 12.7, 10.85, and 8.75 mag, respectively; for the AAVSO $BVRI$ data, these were 15.55, 12.7, 10.85, and 8.75 mag, respectively; for the ASASSN, ATLAS, ZTF values of 13.6, 12.6, and 13.95 mag were subtracted.
Errorbars of the NOT measurements are smaller than the markers. The grey arrow and grey symbols illustrate the change from just before the onset of the dimming to the minimum light, assuming that the event was caused by dust reddening ($A_V$=1.3 mag and $R_V$=7; Appendix \ref{appendix-reddening}) that matches observations nearest to the minimum (in March). }}  
    \label{fig-colors}
\end{figure*}

\section{Photometric evolution during the 2026 dimming}\label{sect-photEvol}
Although our multiband photometry very sparsely covers the 2026 dimming, it implies strong constraints on the nature of the event. Scaled photometric measurements are compared in Fig. \ref{fig-colors}, where the measurements were arbitrarily shifted to the second epoch of NOT observations. The evolution in $BVRI$ bands follows closely the overall evolution seen in the densely spaced ASASSN data. The amplitude in the $U$ band is distinctively shallower than in other bands. {{This evolution can be understood bearing in mind that the photometric magnitudes represent two stars in the V838 Mon system: while the M star dominates in $gVRI$ bands, the contribution of the hot B star is not insignificant in the $U$ band.

The observed photometric evolution is difficult to explain by a temperature change in the photosphere of M star alone, since variations in molecular band opacity would produce color changes markedly different from those observed. Such temperature-triggered opacity changes are a well-known phenomenon in M-type miras \citep{reid} and lead to a visual magnitude change of even 3 mag. This scenario is, however, disfavored for V838 Mon by spectroscopic observations presented below, which imply temperature changes of 100--200 K. 
Even a 200 K temperature change around $T_{\rm eff}\simeq3200$ K would change the bolometric luminosity by only $\sim25\%$ at a fixed radius, substantially less than required to explain the observed optical dimming (cf. Fig. \ref{fig-lightcurve}). The dimming therefore requires an additional mechanism.

Instead, we propose that the M star was partially obscured by a transiting dust cloud that did not cover the hot B star. As described in Appendix~\ref{appendix-reddening}, extinction with $A_V=1.3$ mag and $R_V\gtrsim7$ reproduces the photometric changes between maximum and minimum light well. This is illustrated in Fig. \ref{fig-colors} where we apply the extra reddening to points near NOT epoch 1. The grey symbols represent reddened magnitudes, which match relatively well the observed ones close to the deepest phase of the dimming.

An alternative possibility is occultation of part of the stellar disk by an optically thick body (e.g., a solid object). In Appendix \ref{appendix-reddening} we discuss that, due to insufficient wavelength coverage, the dusty cloud we considered above may be characterized by grey extinction, with $R_V \to \infty$, consistent with such an interpretation. However, we believe that such a partial eclipse is unlikely because similar variability at lower amplitude has been observed in V838 Mon before and is not strictly regular (Sect. \ref{sect-intro}). Additionally, an eclipse would cause a shift in the net velocity of V838 Mon's photospheric features (Rossiter-McLaughlin effect), which is not observed during the dimming in our high-resolution spectra described below.


We favor the interpretation where the M star was obscured by a dusty cloud. The short duration of the event suggests that only a modest dust column was produced along the line of sight and the cloud was not very extended (i.e., was likely smaller than the stellar disk). It then quickly dispersed through expansion or by moving away from the line of sight of the M star. The asymmetry in the slope during dimming and recovery phases may suggest the latter. Using canonical ISM relations between reddening, gas and dust column densities \citep{Allen}, we find that the derived reddening can be converted to a dust column density of 7.8$\cdot$10$^{-5}$ g cm$^{-2}$. Assuming further that half of the stellar disk was covered by the dust puff \citep[as during the Great Dimming;][]{MontargesDimmingNature}, the total mass is 1.3$\cdot$10$^{23}$ g. This is roughly equivalent to half the mass of asteroid Pallas. We caution, however, that it is rather unlikely that the assumptions and conversion factors hold for the warm circumstellar dust.
}}

\section{Deep dimming phase as seen by SALT/HRS}
{{We now turn to spectroscopy to shed more light on the nature of the recent V838 Mon dimming. The initial phase of the dimming was not covered by our spectroscopy, but \citet{Goranskij2026ATel} reported low-resolution spectra acquired on 6 Dec 2025 and 19 Feb 2026 (HJD 24601015 and 24601090) which show V838 Mon unchanged compared to earlier low-resolution observations of this object. Our earliest high-resolution spectra observations are those from SALT, acquired just after the dimming minimum. Below, we discuss all changes that can be traced in the SALT differential spectrum and investigate which of the features can be directly linked to the dimming event.}}

\subsection{Emission lines from the companion}
Our SALT spectrum shows only subtle differences between 2024 and 2026, even though the 2026 spectrum was obtained close to the deepest phase of the dimming. The most striking features in the SALT 2026 spectrum are hydrogen lines exhibiting clear emission components. Among the hydrogen transitions covered, only Ba-H$\alpha$\footnote{Ba-H$\alpha$, H$\beta$, H$\gamma$, etc., designate Balmer transitions. Similar notations are used for Paschen (Pa) and Brackett (Br) lines.} and Ba-H$\beta$ are seen in emission in the differential spectra; that is, both increased in brightness since 2024. Their differential profiles are shown in the top panel of Fig. \ref{fig-emProfilesSALT}.  The rise in fluxes is especially apparent for Ba-H$\beta$, which was virtually absent in 2024. In the differential spectrum, both lines display profiles similar to an inverse P-Cyg shape, with emission centered at the heliocentric velocity of $\approx$25 \kms, the crossover to absorption at 35 \kms, and absorption stretching out to 85 \kms.  Ba-H$\alpha$ in the SALT differential spectrum shows residual emission above 85 \kms (hereafter called red emission component). {{The emission may extend to $\sim200$ \kms, although the TiO-dominated pseudocontinuum prevents us from reliably determining its full velocity extent. No other Balmer lines are apparent in SALT spectra, consistent with Case B decrement ratios.}}


We postulate that these hydrogen features are not directly related to the dimming. In earlier observations of V838 Mon \citep[e.g.][]{Liimets}, Ba-H$\alpha$ appeared as a variable double-peak emission. In a forthcoming paper, it will be shown that this emission arises from around the hot companion, where the wind of the coalesced star is photoionized by the radiation of the hot star as the wind is passing by. The location of the H$\alpha$ emission near the B star has been confirmed by direct H$\alpha$ imaging with VLT/SPHERE (Kamiński et al., in prep.). Long-term monitoring \citep[e.g.,][]{Liimets} has shown that the H$\alpha$ emission has been increasing in intensity for many years, but with some irregular variations within its complex profile, including the one near 25 \kms. We therefore believe that the changes and the emission observed by SALT between 2024 and 2026 are not related to the 2026 dimming event of the M star. Some changes in hydrogen emission are, however, observed in later phases of the recovery from the dimming (see below). 

Narrow lines of [\ion{O}{I}] $\lambda$6300 are observed in both SALT epochs and nearly cancel out in the differential spectrum. The residual emission is centered near 52 \kms, that is, at the center of the absorption component of the differential Ba-H$\alpha$ absorption. The $\lambda$6300 line and a few weak [\ion{Fe}{II}] and \ion{Fe}{II} lines are also formed near the companion and are unrelated to the dimming. These emission lines are compared in the top and bottom panels of Fig. \ref{fig-emProfilesSALT}.

\begin{figure}
    \centering
    \includegraphics[width=1.0\columnwidth,page=2, trim=0 20 0 0, clip=True]{plotSALT-EmLinesDiff-NICE.pdf}    
    \includegraphics[width=1.0\columnwidth,page=9, trim=0 20 0 0, clip=True]{plotSALT-EmLinesDiff-NICE.pdf}
    \includegraphics[width=1.0\columnwidth,page=7, trim=0 0 0 0, clip=True]{plotSALT-EmLinesDiff-NICE.pdf}
    \caption{Emission line profiles in the post-dimming epoch as seen in SALT spectra relative to the 2024 epoch. Dotted vertical lines are drawn at velocities of 23, 55, and 85 \kms, corresponding to the Ba-H$\alpha$ (blue) peak, mid Ba-H$\alpha$ absorption, and red Ba-H$\alpha$ emission peak, respectively. Telluric features were blanked.}
    \label{fig-emProfilesSALT}
\end{figure}

\subsection{Lines from the M-star wind}
Both lines of the \ion{Na}{I} resonance doublet $\lambda\lambda$5889,5895 show subtle changes in the differential spectrum. Over the last two decades, the doublet has displayed a wide, P-Cyg-like, profile \citep[cf.][]{KamiLit}. The SALT 2026-2024 differential spectrum suggests a mild enhancement in the absorption component of this profile and a slight increase in the emission component, in both sodium doublet lines. These long-lived spectral features are most likely related to the post-merger wind of the M star. The variability recorded with the SALT spectra is typical for winds of late-type stars. The observed changes cannot thus be directly linked to the dimming. The optical resonance doublet of \ion{K}{I} has displayed similar profiles to \ion{Na}{I} in the post-outburst spectra of V838 Mon. In the SALT observations, it is strongly contaminated by telluric features and underlying changes in the TiO $\gamma$ (0,1) bands. The less-contaminated $\lambda$7700 line had gained extra emission components between 2024 and 2026, one near 80 \kms\ overlapping with the rising \ion{Na}{I} wind emission, and one at $\approx$+20 \kms\ (see Fig. \ref{fig-emProfilesSALT}). The latter is very likely related to the dimming because this velocity component becomes stronger at later epochs discussed below. 

Similarly, the semiforbidden line of \ion{Ca}{I} $\lambda$6572 with $E_{\rm low}$=0 is an emission line in both SALT spectra and has gained intensity at 88 \kms\ between 2024 and 2026. It may be associated with a weaker blueshifted absorption component, making the feature also  P-Cyg-shaped. It is likely related to the component in the M-star wind that is also seen in \ion{Na}{I} and \ion{K}{I} lines.   

\subsection{Changes in molecular bands}
The strongest positive signal in the differential SALT spectrum is observed for the highly saturated heads of the strongest absorption bands of TiO and VO. The five strongest features belong to TiO $\gamma'$ system: ($v_{\rm up}$,$v_{\rm low}$)=(0,0) at 6216 (R$_3$) and 6223 \AA\ (Q$_3$); and of TiO $\gamma$: (1,0), (2,1), and (0,0) at 6652 (R$_3$), 6709 (R$_1$), and 7055 \AA\ (R$_3$). Changes in the molecular bands occur throughout the covered spectral range, but their relative amplitudes are largest in the low-$v$ bands of the TiO $\gamma$ and $\gamma'$ systems and the VO $B-X$ (0,0) bands. We attribute these changes primarily to a decrease in optical depth. This `softening' of the saturated bandheads has been observed for several years and will be analyzed in a dedicated study. None of these changes can be assigned to the dimming alone. 

Additionally, low-amplitude changes in the molecular spectra, especially far from the heads of bands requiring low excitation, suggest that the molecular photosphere was 100--250 K hotter in 2024 than in the 2026 SALT epoch. This is especially apparent when the SALT spectra are compared to synthetic spectra of M-type stars. We used the PHOENIX/1D NewEra model atmosphere grid \citep{Phoenix2025} with lowest-gravity stars and subsolar metallicity, [Fe/H]=--1, justified by the location of V838 Mon in the Outer Arm. Spectra with effective temperatures near 3200 K explain very well almost the entire SALT/HRS spectral range. To investigate changes of TiO excitation in the photospheric component of the spectrum, we analyzed in particular bands involving high values of $v_{\rm low}$, for instance, the TiO $\gamma$ (0,1) bandhead near 7590 \AA\ and absorption from the TiO $\gamma$ (2,3) in 7740--7840 \AA. The relative intensities and band shapes suggest a $\sim$3300 K photosphere in 2024 and $\sim$3100 K in 2026. Detailed atmosphere modeling would be required to refine these temperature estimates, especially if non-LTE conditions, such as shocks, should be considered. Changes are also seen in the `infrared' \ion{Ca}{II} triplet, but they are difficult to quantify due to the modest quality of the SALT spectra. 

{{To summarize the SALT observations, the spectrum near the minimum light shows no emission lines that can be uniquely assigned to the event, and we find only a modest change in the photospheric TiO temperature.}}

\section{Recovery phase as seen by PEPSI}
\begin{figure}
    \centering
    \includegraphics[width=1.0\columnwidth,page=1]{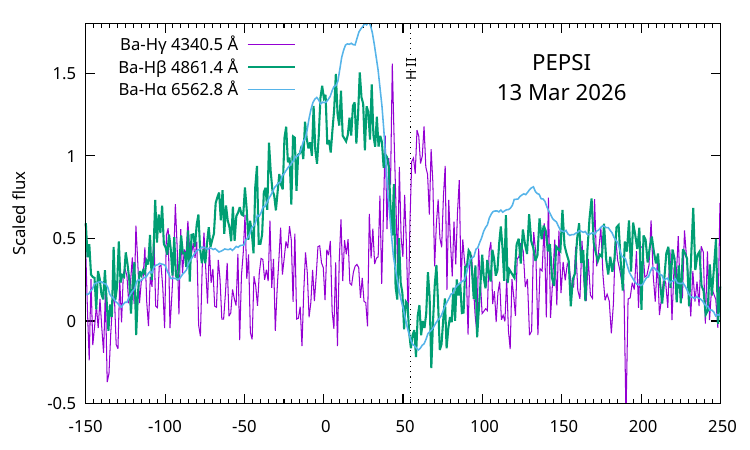}
    \includegraphics[width=1.0\columnwidth,page=2, trim=0 0 0 10]{plotPEPSIemLines.pdf}
    \caption{Sample emission lines observed with PEPSI. Fluxes were arbitrarily scaled. The vertical line is drawn at the central velocity of the \ion{H}{II} region of this epoch, i.e., at +55 \kms.} 
    \label{fig-pepsi}
\end{figure}

The PEPSI spectrum closely resembles the SALT spectrum obtained 10 days earlier. The long-term changes described above continued: the strong low-$v$ bands of TiO and VO soften, but the temperature-sensitive TiO bands from the photosphere show a further increase in temperature over the 10 days, which, however, is smaller than the 100 K step in synthetic spectra we used for comparison. Moreover, emission components of the Ba-H$\alpha$, \ion{Ca}{I}], and \ion{K}{I}  lines decreased in continuum-normalized intensity by $\lesssim$25\%, which is comparable to a change in the ASASSN $g$-band flux level between the two epochs (cf. Fig. \ref{fig-lightcurve}), suggesting these features track the overall continuum rather than an intrinsic line variation. As explained earlier, these features are related to the  \ion{H}{II} region surrounding the B star and the wind surrounding the M star. These components do not vary significantly on the timescale of 10 days. Sample lines from the \ion{H}{II} region are shown in the lower panel of Fig. \ref{fig-pepsi}. They were centered near +55 \kms\ in the PEPSI epoch (Fig. \ref{fig-pepsi}).

The most significant spectroscopic difference relative to the SALT spectrum is observed in the blue wing of Ba-H$\beta$, which increased its continuum-normalized flux by $\approx$30\%. This was associated with weak emission in Ba-H$\delta$ and Ba-H$\gamma$ transitions, even though they are observed at a modest S/N (Ba-H$\gamma$ is seen only in differential, i.e.  PEPSI minus SALT, spectra). This new emission component is slightly redshifted with respect to the \ion{H}{II} region, i.e., to about +60 \kms, as shown in Fig. \ref{fig-pepsi}. As later spectra show, this is a component that can be interpreted as arising from the pulsation shock, and the PEPSI observations constitute the earliest signature of its appearance after the minimum.

\section{End of dimming as seen by Xshooter and UVES}
{{Xshooter and UVES spectra were obtained consecutively on the same night and provide a glimpse of the system at the final phases of the dimming, i.e., when all photometric bands almost recovered their pre-dimming magnitudes.}}

\paragraph{Long-timescale changes}
We first compare our Xshooter spectra from 2026 and 2024 to investigate long-term variations. As in the case of SALT 2026-2024 differential spectra, the most pronounced changes between the 2026 and 2024 Xshooter spectra are seen in the deep molecular absorption bands of low-excitation transitions of TiO and VO. They appear as emission features in the 2026-2024 differential spectrum. These are caused in part by the decreasing optical depth (column density) of the expanding merger ejecta, unrelated to the dimming. 

\subsection{Photosphere changes}
The photospheric component, as in the case of the SALT epochs, can be well reproduced by synthetic spectra of $T_{\rm eff} \approx$3200 K. Comparing the two Xshooter epochs, we found that the photosphere was slightly cooler in 2026 than in 2024, but this time only by about 100 K. While quantifying the temperature on the absolute scale for the spectra is uncertain, we can be quite confident about the difference in the temperature. The Xshooter spectra, having wider spectral coverage and better telluric correction, allowed for a more thorough comparison to 2024 data than the SALT spectrum. For instance, Xshooter spectra also include the TiO $\delta$ bandhead near 8862 \AA\ useful for temperature diagnostics. However, the scaled NIR arm spectra are virtually indistinguishable between the two epochs. Even the CO first-overtone absorption band appears to be unchanged.

\begin{figure*}
    \includegraphics[width=\textwidth]{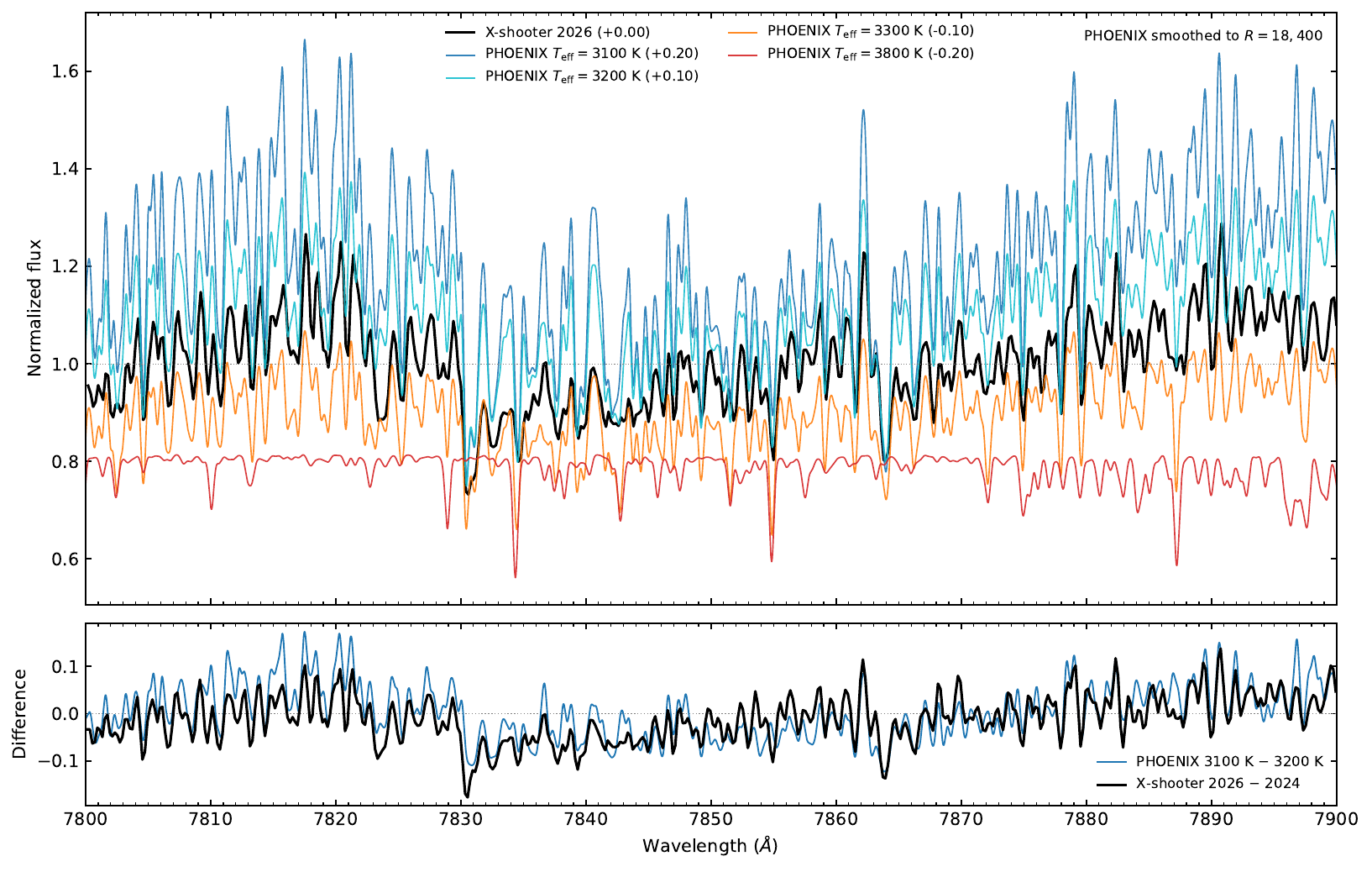}
    \caption{{{Top: The Xshooter spectrum from Apr. 2026 is compared to PHOENIX synthetic spectra near $T_{\rm eff} \approx$3200 K. Bottom: Differential spectrum of Xshooter 2026-2024 compared with the difference in PHOENIX spectra at 3100 and 3200 K. The similarity of the differential spectra indicates that the photospheric temperature changed by only $\sim$100 K between 2024 and 2026.}}}\label{fig-phoenix}
\end{figure*}

Compared to the 2026 SALT spectrum acquired a month earlier and closer to the dimming minimum, the Apr. 2026 Xshooter spectrum shows a marginally hotter photospheric spectrum, by $\lesssim$100 K. {{This is illustrated in Fig. \ref{fig-phoenix}}}. The temperature change is at the limit of the PHOENIX NewEra grid of synthetic spectra we were using as a comparison, which is originally spaced by 100 K. Some higher excitation TiO bands are certainly deeper in the Xshooter 2026 observations.  This suggests that the star was becoming only slightly hotter during the recovery from the dimming. Since the surface of the star closely resembles those of the AGB stars and red supergiants, these temperature differences are smaller than the temperature variations across a stellar disk at any given time \citep{Freytag2024}. 

\subsection{Shock emission lines}
We now focus on the discrete atomic emission features revealed by the 2026-2024 Xshooter differential spectrum. These are listed in Table \ref{tab-xshooter-lines}.  They include hydrogen recombination lines seen across all 3 Xshooter arms and include lines from the Balmer (10 lines), Paschen (3), and Brackett (1) series, most of which were absent in the PEPSI spectrum (if covered). The emission fluxes based on Gaussian fits are listed in Table \ref{tab-xshooter-lines}. 

\begin{figure}
    \centering
    \includegraphics[width=1.0\columnwidth]{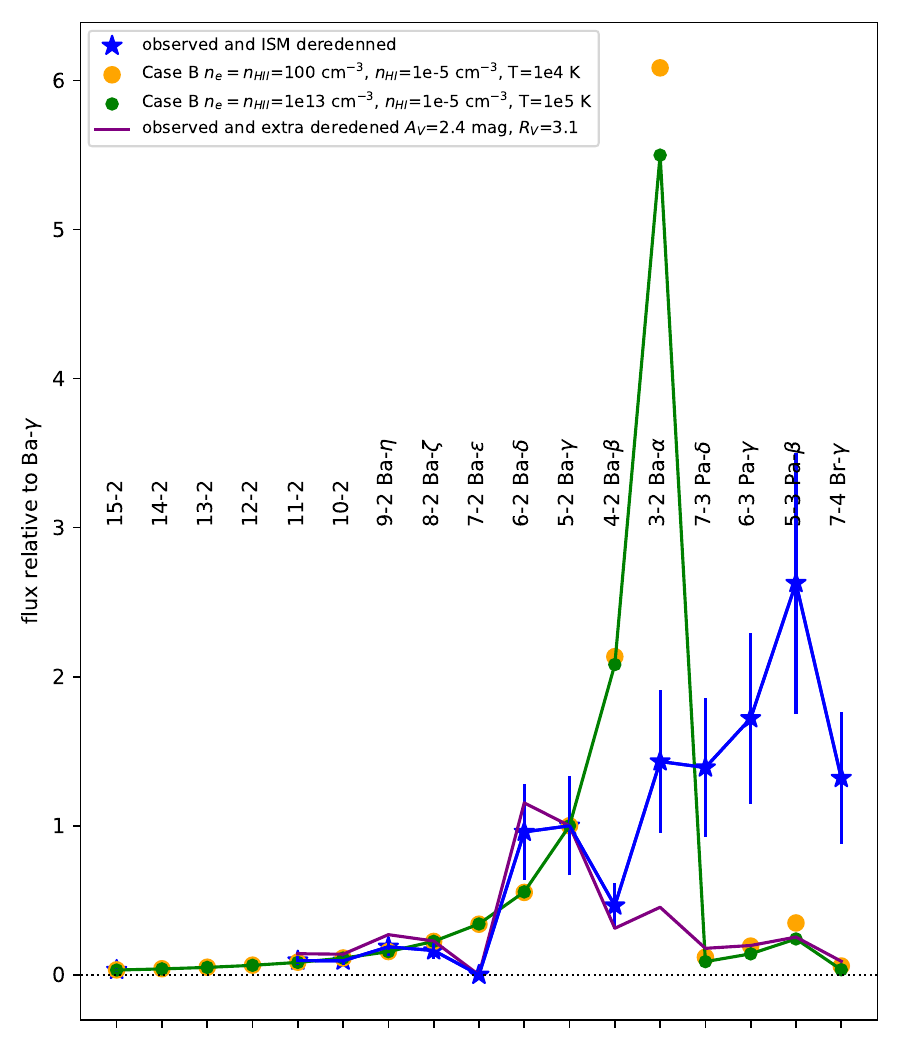}
    \caption{{{Relative intensities of hydrogen emission components seen in Xshooter differential spectrum. Intensities are relative to Ba-$\gamma$. Simulated intensities are shown for two Case B scenarios: a rarified nebula (orange) and an atmospheric Mira-like shock (green). Calculations were made for a classical low-density nebula and for conditions met in photospheric layers of giants \citep[cf.][]{MiraShocks}.  The bluest and reddest observed lines match the Case B scenario if extra extinction is applied, as shown with the purple line.}}}
    \label{fig-Hdecrement}
\end{figure}

\subsubsection{Hydrogen line profiles}
\begin{figure}[h!]
    \centering
    \includegraphics[width=1.0\columnwidth,page=1, trim=0 5 0 10]{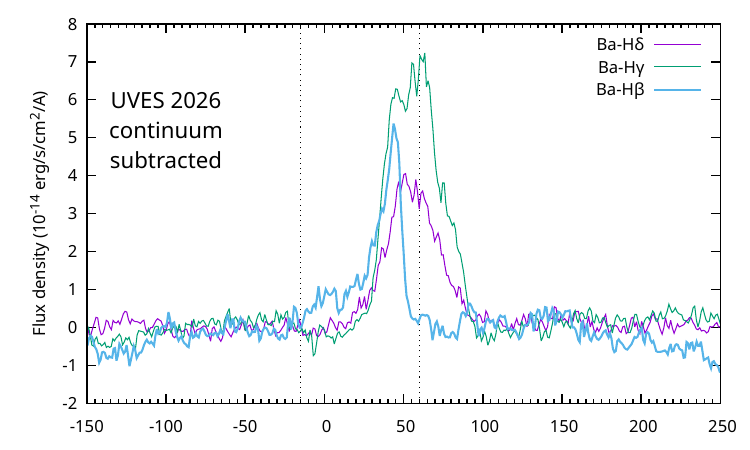}
    \includegraphics[width=1.0\columnwidth,page=1, trim=0 0 0 5]{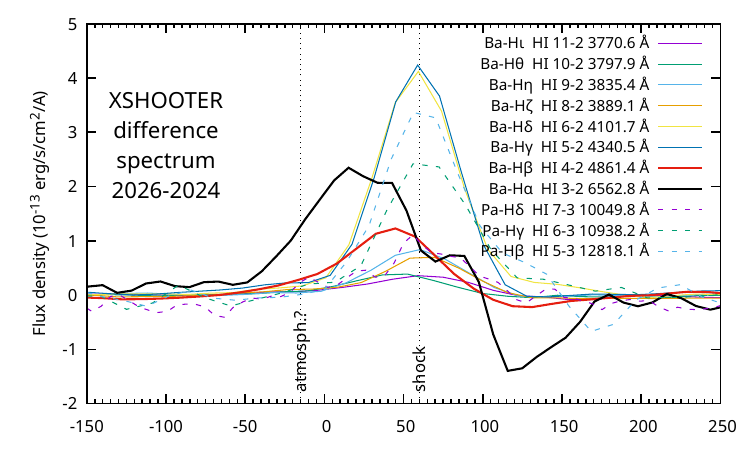}
    \includegraphics[trim=0 15 0 5,width=1.0\columnwidth,page=2, clip=True]{plotXshootEmLinesDiff-NICE2.pdf}
    \includegraphics[trim=0 10 0 5,width=1.0\columnwidth,page=3]{plotXshootEmLinesDiff-NICE2.pdf}
    \caption{Emission line profiles in the post-dimming epoch as seen by UVES (top panel) and Xshooter (middle and bottom panels). The Xshooter spectra were corrected for pseudocontinuum with 2024 Xshooter data. Dotted vertical lines are drawn at --16 and 60 \kms.}
    \label{fig-emProfilesXshUVES}
\end{figure}

We focus first on the Ba-H$\alpha$ and Ba-H$\beta$  transitions. In earlier observations of V838 Mon, Ba-H$\alpha$ appeared as a variable double-peak emission \citep{Liimets} and is most likely related to material around the hot companion. Ba-H$\beta$ from this \ion{H}{II} region was hardly noticeable in spectra from before the 2026 dimming but is detected in the 2026 Xshooter spectrum and is even stronger than in the PEPSI spectrum from 10 days earlier. Both Ba-H$\alpha$ and Ba-H$\beta$ were much stronger during the dimming recovery phase than in any published spectra from the last two decades. Although variable, the emission from the \ion{H}{II} region near the companion was mostly subtracted in the differential Xshooter spectrum we analyze here. However, the differential emission of the two lines is still readily different from all the other hydrogen lines---while most lines in the Xshooter differential spectrum have pure-emission Gaussian profiles centered near a velocity of 60 \kms, Ba-H$\alpha$ and Ba-H$\beta$ display an absorption component at 125 \kms\ and the emission is visibly shifted to the blue compared to other H lines. This profile difference is also visible in UVES spectra obtained at the same epoch but at higher spectral resolution and in the earlier PEPSI spectra (but at a modest S/N). The strongest lines captured with UVES are shown in Fig. \ref{fig-emProfilesXshUVES} (top panel). 
Such distinct profiles of Ba-H$\alpha$ and Ba-H$\beta$, compared with the other Balmer transitions, are commonly observed in hydrogen emission produced by pulsation shocks in Mira variables near maximum light \citep{Joy1947,Gillet,Gillet88}. {{This is illustrated in Appendix Fig.~\ref{fig-RTCyg}, where the Balmer-line profiles of V838 Mon are compared with those of the Mira variable RT Cyg.}}

The remaining Balmer lines are dominated by an emission component narrower than the full emission-plus-absorption profile of Ba-H$\alpha$ (see Fig. \ref{fig-emProfilesXshUVES}). Ba-H$\gamma$ has a full width at half-maximum (FWHM) of 49 \kms\ and a total width of $80\pm10$ \kms. Additionally, the lines observed at UVES resolution have a  `jagged' triple substructure, reminiscent of shock-induced hydrogen lines of miras \citep[cf.][]{Gillet,Alvarez2000,Fabas}{{; see Fig. \ref{fig-RTCyg}.}} The total width of Ba-H$\gamma$ is remarkably similar to those measured in Mira variables: $90\pm10$ \kms\ in $o$ Ceti \citep{Fabas}, $\chi$ Cyg \citep{LopezArtiste}, S Car \citep{Gillet88}, and RT Cyg \citep[][Fig.~\ref{fig-RTCyg}]{Alvarez2000}, and about 80 \kms\ in the six Mira stars studied by \citet{RichterWood}.

Our final note on the hydrogen line profiles in the Xshooter differential spectrum is that Ba-H$\alpha$, H$\beta$, and H$\gamma$ display an absorption component spanning from around 70 to 170 \kms. The Ba-$\alpha$ absorption is the strongest and clearly reaches beyond the velocity range occupied by emission in other hydrogen lines. This suggests a separate cool hydrogen component is present in the system, likely well above the atmosphere, but on the line of sight of the M star. 

\subsubsection{Hydrogen line ratios}
The strengthening of hydrogen recombination lines in April 2026, i.e. at the end of the dimming, is most interesting. Compared to the SALT spectrum obtained a month earlier, Ba-H$\alpha$, scaled to the local continuum, is actually weaker in the later epoch, whereas all other lines strengthened. The hydrogen emission lines show rather peculiar line ratios. In particular, the Balmer series does not follow the canonical Case B decrement, as shown in Fig. \ref{fig-Hdecrement}. 


{{Figure~\ref{fig-Hdecrement} compares the observed line ratios with two Case B calculations from HyLight \citep{HyLight}, representative of a low-density nebula and the high-density photospheric conditions of a Mira star. The predicted ratios are very similar for most transitions, whereas the observed V838 Mon ratios show clear departures from Case B. Excluding Ba-H$\alpha$, Ba-H$\beta$, and Ba-H$\epsilon$, the remaining ratios can be reconciled with Case B by introducing additional extinction. A least-squares fit with the CCM89 law and $R_V=3.1$ yields $A_V=2.4$ mag. When both parameters are free, the fit gives $A_V=4.9$ mag and favors increasingly large $R_V$, formally tending toward $R_V\rightarrow\infty$.}}

Even with the extra reddening Ba-H$\alpha$, Ba-H$\beta$, Ba-H$\epsilon$ are too weak. While Ba-H$\alpha$ and Ba-H$\beta$ can be affected by the \ion{H}{II} region of the companion, Ba-H$\epsilon$ should not be affected at all. Ba-H$\epsilon$ overlaps with the wing of the strong \ion{Ca}{II} H line, shows nearly no net emission, and rather appears as pure absorption in the differential spectrum. Similar `anomalies' in the three Balmer lines are common in pulsating M-type miras near maximum light \citep[cf.][]{Yao2017}. Such hydrogen-line behavior has been partially reproduced theoretically as emission from structured pulsation shocks \citep{FadayevGillet2004}. As described in \cite{CastelazLuttermoser} and \cite{Siviero}, the radiative transfer of photons arising in the sub-photospheric pulsation shock of miras is strongly affected by overlapping features of TiO and the \ion{Ca}{II} H line, whose carriers are located in a screen between us and the shock. This similarity between the observed post-dimming emission of V838 Mon and pulsating M-type miras indicates that hydrogen emission comes from within the deep photospheric layers of the atmosphere \citep[cf.][]{MiraShocks,Fadeev2001,FadayevGillet2004}. The lack of low-density shock tracers, such as the optical nebular lines, also suggests a high-density environment of the hydrogen emission region. The relatively weak and narrow hydrogen emission can be interpreted as the shock covering only a small fraction of the stellar disk. This is often the case for these photospheres affected by huge convection cells \citep{Freytag2024,ChiavassaRev2024}.

{{The additional extinction required to bring the hydrogen-line ratios close to Case B is potentially important for understanding the V838 Mon system as a whole. We interpret it as evidence for an additional local, predominantly circumstellar, extinction component toward the M star, or possibly toward both stars. This extinction need not be directly associated with the transient dust cloud proposed in Sect.~\ref{sect-photEvol}, because the Xshooter spectrum was obtained near the end of the recovery phase, when the extinction produced by that cloud should already have been small. One possibility is that repeated dimming events have gradually accumulated dust along the line of sight. Interestingly, both independent extinction estimates---that derived from the photometric dimming and that inferred from the hydrogen-line ratios---favor large values of $R_V$, formally approaching $R_V\rightarrow\infty$ in some fits, and thus nearly grey extinction. Nevertheless, we favor the interpretation that most of the additional extinction inferred from the hydrogen decrement predates the 2002 eruption and is related to earlier mass loss in the V838 Mon system. Such a persistent circumstellar component could also explain the long-standing problem that the progenitor of V838 Mon and its B-type companion appeared $\sim2$ mag fainter than other B-type members of the V838 Mon cluster \citep{AfsarBond,TylendaKaminskiEcho}.}}

\subsection{Other emission features}

Firmly-identified emission lines, besides those of hydrogen, are surprisingly few in the Xshooter differential spectrum. The strongest emission is from the resonance doublet of \ion{K}{I} at $\lambda\lambda$7664, 7698, followed by three resonance lines of \ion{Cr}{I} between 4254--4290 \AA, and 3--4 other lines again of \ion{Cr}{I} near 5205 \AA\ but from excited levels at 7593 cm$^{-1}$ ($\approx$10\,900 K); in the NIR spectrum, we also find a blend of 3 lines of \ion{S}{I} near 11\,343 \AA, but their identification is uncertain. Sample lines are shown in the two lower panels of Fig. \ref{fig-emProfilesXshUVES}. Surprisingly, no differential emission is seen in the resonance lines of \ion{Na}{I}, \ion{Li}{I}, \ion{Rb}{I}, and \ion{Mn}{I}, which sustained an emission component long after the 2002 eruption. 

As shown in Fig. \ref{fig-emProfilesXshUVES}, the emission of resonance transitions is shifted to a central velocity of --16 \kms. It is thus blueshifted by about 75 \kms\ relative to the center of the hydrogen lines (apart from Br-H$\alpha$ and H$\beta$); and is blueshifted by 90 \kms\ relative to the center of the \ion{Na}{I} wind features observed earlier in the dimming with SALT. Assuming this gas was ejected from the M star, its radial velocity relative to the stars would be $\approx-90$ \kms. As a working hypothesis, we propose that this is a localized (i.e., of limited volume) atmospheric response to the shock braking from sub-photospheric layers. This gas is of low-ionization and may be filling a small part of the densest layers of the wind (as no low-density tracers such as forbidden lines are observed). The lack of velocity components at or redshifted relative to the systemic velocity then means that the parcel of gas is located close to the near side of the stellar disk and is hotter than the photosphere itself (to produce net emission against the stellar disk). Its temperature is thus likely a few 1000 K  and definitely much lower than 10\,000 K.

In Mira stars, lines of \ion{Fe}{I}, \ion{Fe}{II}, [\ion{Fe}{II}] have been successfully used as a diagnostic tool of the shocked gas \citep{RichterWoodII,Fokin}. These lines flare up shortly after the Balmer transitions reach maximum intensities \citep{RichterWood}. We searched for the useful iron lines in our spectra and differential spectra. Omitting lines that most likely come from the \ion{H}{II} region, we only find 5 weak lines that may belong to \ion{Fe}{I}: at 3906, 4427.3, 4465.65, 14\,233 and 15\,757 \AA. No lines of ionized iron were found. Except for the IR lines, which require very high excitation, these transitions are from levels 22\,000--26\,000 cm$^{-1}$ above the ground. They therefore require higher temperatures than the chromium and potassium lines discussed above. These \ion{Fe}{I} lines are all centered near velocity 30$\pm$15 \kms, which is intermediate between the low-excitation chromium-bearing gas and the hot hydrogen-emitting plasma. There are so many iron transitions that a chance alignment between transitions cannot be excluded. If correctly identified, however, these lines may represent regions of the atmosphere more strongly excited by the shock, and thus closer to the photosphere, than the gas bright in the \ion{Cr}{I} lines. Two of the tentatively identified iron lines (at 4427.3 and 4465.65 \AA) were targeted in a dedicated monitoring study of Mira stars of \citet{RichterWood}, but were found in one source only, so no meaningful comparison is possible.

\section{Discussion}

We interpret the presented observations as evidence of a pulsation shock in the merger remnant V838 Mon. Analogous pulsations result in mass loss and atmospheric shocks in RSGs and Mira stars. We propose that the dimming event with the minimum in 2026 was directly caused by the previous pulsation shock, which was not observed spectroscopically (i.e., during the second half of the region marked as `dimming 2025' in Fig. \ref{fig-lightcurve}), when the star was a daytime object. This shock either produced a clump of dust that crossed our line of sight or, alternatively, caused a drastic dimming of a large fraction of the stellar disk at nearly constant net effective temperature. These two scenarios are discussed in more detail below. 
After the minimum, a new pulsation shock became apparent through the hydrogen recombination lines, which we first detected in the PEPSI spectrum about a month after the $g$-band minimum. The shock signatures strengthened during the recovery phase. Radiation from the shock, traced by the strengthening hydrogen emission, photo-excited gas in the overlying atmospheric layers, giving rise to low-excitation metal emission centered near a radial velocity of --16 \kms. Thus, the hydrogen- and metal-emitting regions need not be spatially coincident or in direct contact.
If our interpretation is correct, the outwardly propagating disturbance may eventually trigger a subsequent dimming event.

\subsection{Dimming mechanism}
Although in Sect. \ref{sect-photEvol} we advocated a scenario where the dimming is caused by a transpassing clump of freshly coagulated dust, other interpretations are possible. Even in the case of the Great Dimming of Betelgeuse, much better covered by observations than the event in V838 Mon, there is an ongoing controversy about which physical processes cause the deep dimming events. A transiting dust cloud proposed for Betelgeuse \citep[e.g.,][]{Dupree2020,Dupree2022,LevesqueMassay2020,MontargesDimmingNature} and other objects \citep[e.g.,][]{dimmingRWCep,KamiVYclumps} naturally explains the clumpy winds of these stars and the erratic visual light curves (since only a small fraction of the clumps are ejected on the line of sight toward the star). However, the lack of corresponding dust emission at longer wavelengths \citep{Gehrz2020,Dharma} and signatures of decreasing temperature \citep{Harper2020} led to an alternative scenario, which is linked to the actual dimming of a large fraction of the stellar disk. Because this scenario can be applied to V838 Mon, we recapitulate its main points below. The $V$-band amplitude of the Great Dimming was similar to that in V838 Mon, $\approx$1 mag.

 
Based on direct imaging of Betelgeuse during the dimming, \cite{MontargesDimmingNature} proposed that a large fraction of the stellar surface, up to 79\%, attained lower temperature. Since the low-intensity regions with lower temperatures contribute less to the net flux from the photosphere, temperature determinations from spatially unresolved spectra overestimate the temperature and are not very sensitive to temperature variations. \citet{Harper2020} showed that temperature changes of $\gtrsim$250 K can explain the visual dimming and the modest change in the appearance of the net photospheric spectrum. While the photospheric temperature of V838 Mon (3300 K) is lower than that of Betelgeuse (3650 K), the argument can be repeated for the young merger product and is consistent with the temperature variations of $\lesssim$200 K we derived. Similarly, \cite{MontargesDimmingNature} required only a 300 K change in the photospheric temperature within the `cool hemisphere' to effectively dim the stellar surface in visual bands. They also note that this affects the circumstellar matter by lowering the equilibrium temperature by a few 100 K, which may trigger more effective dust formation, naturally joining the two competing scenarios. Even this model, however, overpredicts the flux change in the NIR bands. We believe that V838 Mon has undergone such a hybrid dimming because we found evidence for modest temperature change and for extra extinction in the deep phase of the dimming.


The Great Dimming of Betelgeuse is widely interpreted as involving photospheric motions triggered by convection and pulsations leading to Rayleigh-Taylor instabilities \citep{Harper2020,Freytag2024,2024A&A...685A.124J}.  We suggest that analogous processes operate in V838 Mon. It may even relate to the extreme mass loss phenomena in pre-supernova phases of RSGs \citep{Heger1997,Sengupta2026}.


{The time between the onsets of the two 2025--2026 dimming events is $P\approx220$ d (cf. Fig.~\ref{fig-lightcurve}), somewhat shorter than the 360 d period initially proposed by \citet{goran2020}. Periods of 200--400 d are typical of both Mira variables and low-mass RSGs. To within a factor of a few, the recent variability of V838 Mon follows empirical period relations for convective giants, such as $P\propto L/M$ \citep{Heger1997}. However, for dimming events caused solely by dust clumps originating from random locations on the surface, one would not expect any strong periodicity. Future variability of V838 Mon may clarify the origin of the observed timescales.

\subsection{Shock braking in the photosphere}
The appearance of hydrogen recombination lines just after the minimum and their rise in intensity as V838 Mon climbed back to nominal (or maximal) flux levels is reminiscent of a similar phenomenon in Mira stars. The light variations in miras, unlike the dimming of V838 Mon and Betelgeuse, are mainly caused by opacity variations in molecular bands \citep{reid}. For the recently analyzed dimming events of cool supergiants, we did not find reports of flaring in hydrogen recombination lines. V838 Mon is therefore not a perfect analog of either class of objects in this respect. 
The atmospheric extension of V838 Mon, as measured by the ratio of the photospheric pressure scale height to radius, $H_p/R \approx 0.64\%$, is comparable to that of red supergiants (0.64\% for Betelgeuse) and about four times smaller than in typical Miras stars ($\sim$2.8\%). These are first-order hydrostatic values, though. 
Nevertheless, its hydrogen-line spectrum strongly resembles that produced by pulsation shocks in Mira variables. The recent quasi-period of V838 Mon, presumably associated with pulsation, is also comparable to the observed periods of both Mira variables and red supergiants, although their dynamical timescales differ substantially. V838 Mon thus occupies an interesting intermediate regime, with a radius comparable to those of Mira variables but a luminosity characteristic of RSGs. The radiative transfer of shock emission emerging from sub-photospheric layers is therefore likely to differ substantially among these objects.





Applying the shock models of \citet{FadayevGillet2004} for a shock wave propagation in a 3000 K star, the radial velocity shift of Ba-H$\gamma$ emission of $V_H$=--15 \kms\ relative to the star implies a shock velocity of roughly $2 \times V_H$=30 \kms\ and a Mach number of $\sim$6. This and the measured value $FWHM$ of 49 \kms\ further imply a local density of 10$^{-9}$ g cm$^{-3}$ ($n_H =10^{15}$ cm$^{-3}$) and a shock compression ratio of 12--15 for the emission observed in the UVES epoch. These calculations ignored projection effects and the presence of the cool atmosphere, chromosphere, and extended wind which may affect the profiles by extra absorption. Nevertheless, the model quite satisfactorily explains the V838 Mon observations, adding evidence for pulsation shocks in the atmosphere of the merger remnant.

Unlike Betelgeuse \citep{Dupree2020,Mittag}, V838 Mon did not show any distinguishable features of chromospheric activity related to its dimming. While the \ion{Ca}{II} H and K lines are covered by our spectra, they do not show changes in the wings that go beyond what can be expected from temperature changing by 100--200 K. Future ultraviolet observations of chromospheric activity in \ion{Mg}{II} lines, such as those obtained for Betelgeuse around the Great Dimming \citep{Dupree2020}, should be a better probe of whether the merger remnant has a chromosphere and whether it is excited by the newly discovered pulsations. However, the lack of \ion{Fe}{I} fluorescent emission lines at 4307.90, 4202.03 and 3852.57 \AA\ , which are thought to be pumped by the \ion{Mg}{II} $\lambda\lambda$ 2795, 2802 transitions \citep[][and refs. therein]{RichterWood}, suggests that there was not significant activity in these UV lines during the recovery phase of the 2026 event.

\subsection{Future behaviour}
By analogy with the Great Dimming described by \citet{2024A&A...685A.124J}, the 2026 dimming of V838 Mon may have been triggered by a pulsation shock immediately preceding the event. This precursor shock likely occurred during the recovery from the shallower 2025 dimming episode marked in Fig.~\ref{fig-lightcurve}. If this interpretation is correct, the shock revealed by the hydrogen emission in April 2026 may trigger another dimming event, beginning around mid-2026 and reaching minimum light toward the end of the year. Future monitoring will reveal whether the next dimming event is even more pronounced, or whether the light curve instead returns to lower-amplitude variations as observed for Betelgeuse following its Great Dimming. Given the unusual nature of V838 Mon, still thermally relaxing after the merger, another extreme dimming could indicate a more severe instability and thus warrants detailed observations.

{{
\section{Summary and conclusions}

The 2025--2026 photometric event in V838 Mon closely resembles the deep dimming episodes observed in red supergiants. Its wavelength dependence can be reproduced by additional circumstellar extinction of $A_V\simeq1.3$--1.4 mag, with the available data favoring large $R_V$ and therefore nearly grey extinction. However, the extinction law remains poorly constrained. In particular, we lack sufficiently constraining $U$-band measurements and contemporaneous $JHK_s$ and mid- or far-infrared observations around minimum light, which would provide much stronger constraints on the grain properties and on possible thermal emission from newly formed dust. Whether the dimming was indeed produced by nearly grey extinction should therefore be tested if additional photometry from the minimum becomes available. The close phenomenological similarity to dimming events in red supergiants nevertheless suggests that related atmospheric and circumstellar processes may operate in V838 Mon.

Our spectroscopy covers only the recovery from the dimming minimum. The photospheric spectrum remained remarkably similar to that observed in 2024, with inferred temperature differences of only $\sim$100--200 K, close to the limit that can be established from comparison with the available synthetic spectra. More unexpectedly, the most conspicuous new spectral features did not appear in the spectrum obtained closest to minimum light, but developed during the subsequent recovery. Hydrogen recombination lines from the Balmer, Paschen, and Brackett series strengthened progressively, and their profiles and relative intensities depart strongly from those expected for a classical recombination nebula. Instead, both the line profiles and the peculiar Balmer-line ratios closely resemble those observed in Mira variables near maximum light.

This resemblance provides the strongest evidence that the hydrogen emission arises from a pulsation shock propagating through the atmosphere of the merger remnant. Hydrogen emission associated with Mira pulsation shocks has a highly characteristic combination of line profiles, widths, and anomalous decrements, and we are not aware of another astrophysical phenomenon that provides a comparably close analogue to the spectrum observed in V838 Mon. The appearance of such signatures only 24 yr after the stellar merger therefore makes a compelling case that the merger product has developed strong pulsational instability. In this interpretation, the earlier dimming may itself have been triggered by a preceding pulsation-driven atmospheric disturbance, linking the photometric and spectroscopic phenomena.

The hydrogen-line ratios also reveal an additional result that is largely independent of the transient dimming. Bringing most of the observed hydrogen lines close to Case B ratios requires extra extinction along the line of sight. The fits again favor large $R_V$, suggesting nearly grey circumstellar extinction. This additional component may have accumulated through repeated dust-producing events, although we favor an origin predating the 2002 eruption and related to earlier mass loss from the system. Such persistent circumstellar extinction could naturally explain the long-standing problem that the V838 Mon progenitor and its B-type companion appeared $\sim2$ mag fainter than other B-type members of the surrounding cluster. In addition, the appearance of low-ionization metal emission blueshifted by about 90 \kms\ relative to the stellar rest frame provides evidence that the pulsational disturbance affected material above the photosphere, possibly through radiative excitation by the shock.}}

\begin{acknowledgements} 
T.K. acknowledges funding from grant SONATA BIS no. 2018/30/E/ST9/00398 from the Polish National Science Center. Based on observations made with ESO Telescopes at the Paranal Observatory under programs 109.233N and 116.2ASQ. This research has made use of NASA's Astrophysics Data System. 
Some observations reported in this paper were obtained with the Southern African Large Telescope (SALT) under programs 2024-2-SCI-006 and 2025-2-SCI-008. Polish participation in SALT is funded by MNiSW grant No. 2026/WK/02.
The LBT is an international collaboration among institutions in the United States and Europe. At the time data were acquired for this research, LBT Corporation Members were the University of Arizona on behalf of the Arizona Board of Regents; Istituto Nazionale di Astrofisica, Italy; and The Ohio State University, representing The Ohio State University, University of Notre Dame, University of Minnesota, and University of Virginia.  This research used the facilities of the Italian Center for Astronomical Archives (IA2) operated by INAF at the Astronomical Observatory of Trieste.  Observations have benefited from the use of ALTA Center (alta.arcetri.inaf.it) forecasts performed with the Astro-Meso-Nh model. Initialization data of the ALTA automatic forecast system come from the General Circulation Model (HRES) of the European Centre for Medium Range Weather Forecasts.
Based on observations made with the Nordic Optical Telescope, owned in collaboration by the University of Turku and Aarhus University, and operated jointly by Aarhus University, the University of Turku and the University of Oslo, representing Denmark, Finland and Norway, the University of Iceland and Stockholm University at the Observatorio del Roque de los Muchachos, La Palma, Spain, of the Instituto de Astrofisica de Canarias.The data presented here were obtained with ALFOSC, which is provided by the Instituto de Astrofisica de Andalucia (IAA) under a joint agreement with the University of Copenhagen and NOT.

The ZTF forced-photometry service was funded under the Heising-Simons Foundation grant
no. 12540303 (PI: Graham).

This research has made use of the SVO Filter Profile Service "Carlos Rodrigo", funded by MCIN/AEI/10.13039/501100011033/ through grant PID2023-146210NB-I00. 

This work also made use of Python libraries: Astropy, NumPy, extinction, SciPy, Matplotlib, and synthetic\_photometry.


This work also made use of the GAIA software. GAIA is a derivative of the Skycat catalogue and image display tool, developed as part of the VLT project at ESO. Skycat and GAIA are free software under the terms of the GNU copyright. The 3D facilities in GAIA use the VTK library.

This work has made use of data from the Asteroid Terrestrial-impact Last Alert System (ATLAS) project. ATLAS is primarily funded to search for near-Earth asteroids through NASA grants NN12AR55G, 80NSSC18K0284, and 80NSSC18K1575; by-products of the NEO search include images and catalogs from the survey area. The ATLAS science products have been made possible through contributions from the University of Hawai‘i Institute for Astronomy, Queen’s University Belfast, the Space Telescope Science Institute, the South African Astronomical Observatory, and the Millennium Institute of Astrophysics, Chile.

Based on spectral data retrieved from the ELODIE archive at Observatoire de Haute-Provence (OHP)

We gratefully acknowledge the contributions of the AAVSO observer community, whose photometric data and metadata resources were used in this study and made available through the AAVSO’s scientific archives. 
\end{acknowledgements}

\bibliographystyle{aa}
\bibliography{0bibv838.bib}

\begin{appendix}

\section{Supplementary tables}

\begin{table}
\caption{Sensitivities reached in our spectral observations.}
\label{tab-sn}
\centering
\begin{tabular}{ll c c}
\hline\hline
Date &Instrument & S/N & S/N\\
     &           & \@4200 \AA &\@7000 \AA \\
\hline
15-11-2024 & SALT/HRS & 6 & 162\\
02-03-2026 & SALT/HRS & 4&  179\\
13-03-2026 & LBT/PEPSI & 3& 61\\
02-03-2024 & VLT/Xshooter & 78 & 460\\
10-04-2026 & VLT/Xshooter & 57 & 884\\
10/12-04-2026  &  VLT/UVES & 13 & 27\\
\hline
\end{tabular}
\end{table}

\begin{table*}[htbp]
\centering
\caption{Measurements for emission lines in the Xshooter 2026-2024 differential spectrum.}
\label{tab-xshooter-lines}
\resizebox{\textwidth}{!}{%
\begin{tabular}{lllllllll}
\hline
\textbf{Center} & \textbf{Cent. Err} & \textbf{FWHM} & \textbf{FWHM Err} & \textbf{Flux} & \textbf{Flux Err} & \textbf{Fit RMS} & \textbf{Lab $\lambda_{\rm air}$} & \textbf{ID} \\
(\AA)&(\AA)&(\AA)&(\AA)&(erg/s/cm$^2$/\AA)&(erg/s/cm$^2$/\AA)&(erg/s/cm$^2$/\AA)&(\AA)&\\
\hline
3712.74  & 0.06 & 1.08 & 0.15 & $1.21\times10^{-14}$ & $2.23\times10^{-15}$ & $2.07\times10^{-15}$ & 3711.98  & HI \\
3771.35  & 0.02 & 0.93 & 0.05 & $3.53\times10^{-14}$ & $2.64\times10^{-15}$ & $2.64\times10^{-15}$ & 3770.63  & HI \\
3798.46  & 0.02 & 0.84 & 0.05 & $3.50\times10^{-14}$ & $2.67\times10^{-15}$ & $2.80\times10^{-15}$ & 3797.91  & HI \\
3836.14  & 0.01 & 0.79 & 0.02 & $6.95\times10^{-14}$ & $2.54\times10^{-15}$ & $2.76\times10^{-15}$ & 3835.40  & HI \\
3889.87  & 0.01 & 0.80 & 0.03 & $6.05\times10^{-14}$ & $3.07\times10^{-15}$ & $3.30\times10^{-15}$ & 3889.06  & HI \\
3906.45  & 0.05 & 0.74 & 0.11 & $7.52\times10^{-15}$ & $1.54\times10^{-15}$ & $1.70\times10^{-15}$ & 3906.48  & FeI? \\
4102.53  & 0.01 & 0.81 & 0.01 & $3.55\times10^{-13}$ & $3.58\times10^{-15}$ & $3.84\times10^{-15}$ & 4101.73  & HI \\
4254.11  & 0.03 & 0.86 & 0.06 & $2.88\times10^{-14}$ & $2.78\times10^{-15}$ & $2.88\times10^{-15}$ & 4254.35  & CrI \\
4274.55  & 0.02 & 1.01 & 0.06 & $3.40\times10^{-14}$ & $2.47\times10^{-15}$ & $2.37\times10^{-15}$ & 4274.81  & CrI \\
4289.35  & 0.04 & 0.98 & 0.08 & $2.72\times10^{-14}$ & $3.07\times10^{-15}$ & $2.98\times10^{-15}$ & 4289.73  & CrI \\
4341.33  & 0.01 & 0.82 & 0.01 & $3.70\times10^{-13}$ & $4.89\times10^{-15}$ & $5.23\times10^{-15}$ & 4340.47  & HI \\
4427.63	 & 0.04	& 0.50 & 0.09 &	$6.36\times10^{-15}$ & $1.43\times10^{-15}$	& $1.87\times10^{-15}$ & 4427.31  & FeI \\
4462.24  & 0.01	& 0.28 & 0.01 &	$1.14\times10^{-14}$ & $7.19\times10^{-16}$ & $7.98\times10^{-16}$ & 4461.65  & FeI \\
4861.99  & 0.02 & 1.24 & 0.05 & $1.72\times10^{-13}$ & $9.41\times10^{-15}$ & $8.12\times10^{-15}$ & 4861.35  & HI \\
5205.67  & 0.15 & 5.59 & 0.36 & $3.66\times10^{-13}$ & $3.15\times10^{-14}$ & $1.29\times10^{-14}$ & 5205.00  & CrI-blend \\
6563.22  & 0.06 & 1.94 & 0.13 & $5.29\times10^{-13}$ & $4.76\times10^{-14}$ & $3.30\times10^{-14}$ & 6562.79  & HI \\
7664.46  & 0.05 & 1.66 & 0.12 & $7.89\times10^{-13}$ & $7.78\times10^{-14}$ & $5.79\times10^{-14}$ & 7664.90  & KI \\
7698.38  & 0.04 & 2.17 & 0.10 & $9.05\times10^{-13}$ & $5.53\times10^{-14}$ & $3.60\times10^{-14}$ & 7698.96  & KI \\
10051.77 & 0.20 & 4.53 & 0.48 & $5.14\times10^{-13}$ & $7.17\times10^{-14}$ & $1.87\times10^{-14}$ & 10049.80 & HI \\
10940.57 & 0.05 & 2.40 & 0.12 & $6.45\times10^{-13}$ & $4.34\times10^{-14}$ & $1.55\times10^{-14}$ & 10941.22 & CrI \\
11343.43 & 0.12 & 3.04 & 0.27 & $2.23\times10^{-12}$ & $2.64\times10^{-13}$ & $8.43\times10^{-14}$ & 11343.19 & SI-blend \\
10940.57 & 0.06 & 2.38 & 0.13 & $6.37\times10^{-13}$ & $4.63\times10^{-14}$ & $1.68\times10^{-14}$ & 10938.17 & HI \\
12820.77 & 0.04 & 2.71 & 0.10 & $9.72\times10^{-13}$ & $4.93\times10^{-14}$ & $1.66\times10^{-14}$ & 12818.07 & HI \\
15759.13 & 0.09 & 1.67 & 0.21 & $1.74\times10^{-13}$ & $2.85\times10^{-14}$ & $1.21\times10^{-14}$ & 15757.08 & FeI \\
21664.80 & 0.69 & 19.04 & 1.63 & $4.88\times10^{-13}$ & $5.52\times10^{-14}$ & $7.07\times10^{-15}$ & 21661.18 & HI \\
\hline
\end{tabular}%
}
\end{table*}
\section{Extinction by circumstellar dust}\label{appendix-reddening}
To test the hypothesis that the 2026 dimming was caused by obscuration by circumstellar dust, we estimated the basic extinction parameters, namely the visual extinction, $A_V$, and the total-to-selective extinction ratio, $R_V$. We compared the available photometry near the onset of the dimming, close to the epoch of our first NOT observations, with that obtained around minimum light. Where measurements were not available at exactly the required epochs, we either interpolated the light curves or adopted the nearest representative measurement. For example, for the $B$-band minimum, we adopted the AAVSO measurement at HJD 246173 (cf. Fig. \ref{fig-colors}); despite its relatively large uncertainty, this point follows the overall trend into and out of the minimum. The AAVSO $I$-band data provide good coverage of the minimum but lack a suitable reference measurement immediately before the dimming. Moreover, the NOT $I$-band magnitudes do not appear to be fully consistent with the AAVSO light curve. We therefore adopted an AAVSO measurement obtained near the 2025 maximum as the reference value and assigned it a relatively large uncertainty. The photometry was not corrected for interstellar or any other constant circumstellar extinction, as such corrections are irrelevant for the present analysis, which relies solely on the relative change in brightness in each band during the dimming.

The resulting magnitude differences were compared with a grid of extinction values calculated using the Python package {\tt extinction} and the canonical CCM89 extinction law \citep{CCM89}. This approach evaluates the monochromatic extinction at a specified wavelength. For each photometric band, we adopted the effective wavelength calculated from the corresponding filter transmission curve and our 2026 X-shooter spectrum of V838 Mon prior to dereddening. We then performed a weighted least-squares fit, using inverse-variance weights, to determine the optimal values of $A_V$ and $R_V$ from the combined AAVSO $BVI$, ATLAS $c$, ZTF $g$, and ASASSN $g$ photometry. The optical bands primarily constrain $A_V$, for which the fit yielded $A_V=1.4$ mag, with typical residuals of approximately 0.1 mag. The largest discrepancy between the calculated and observed changes occurs in the ATLAS $c$ band. This may be related to the exceptionally broad wavelength coverage of the $c$ filter, which is not fully represented by our monochromatic treatment.

The $R_V$ parameter is much less well constrained. For the CCM89 law, the fit does not converge to a unique value and instead favors increasingly large $R_V$, formally tending toward $R_V\rightarrow\infty$. The absence of $U$-band measurements at minimum light and the poor $B$-band coverage provide little leverage on $R_V$. In contrast, when adopting the extinction law of \citet{Calzetti}, the fit converges to $A_V=1.3$ mag and $R_V=6.7$. This solution is shown in Fig. \ref{fig-colors}, although it is representative of a broader family of solutions with $R_V\gtrsim5$, including those obtained with CCM89. Moreover, by reducing the fit to the four most reliable measurement sets, we obtain $A_V=1.3$ mag and $R_V=7$ with CCM89. The high-$R_V$ solutions reproduce the observed changes with residuals $\lesssim0.15$ mag, with the largest discrepancy occurring in the AAVSO $I$-band data, for which the pre-dimming reference magnitude is particularly uncertain. Overall, the agreement between the calculated and observed photometric changes is satisfactory, especially given the simplified monochromatic treatment of the extinction.  


Our NOT observations in the $JHK_s$ bands show no change in photometric fluxes relative to 2025 (Sect. \ref{sect-obs-phot}). This is not particularly constraining for the properties of the hypothesized dust cloud, because the observations were done after the dimming was over. Our reddening estimates $A_V=1.3$ mag and $R_V=6.7$ yield extinction values $\Delta J$=0.46, $\Delta H$=0.29, and $\Delta K_s$=0.19 mag. As the dust is expected to have temperatures below $\sim$1200 K, observations at the mid-infrared would be needed to test whether this warm dust caused any excess emission at longer wavelengths. That emission would be visible longer than the dimming phase.

\section{Profile comparison to a Mira}
To illustrate the close resemblance of V838 Mon's Balmer emission lines observed during the dimming to those observed in M-type Mira variables, we extracted spectra of RT Cyg reported in \citet{Alvarez2000}. The spectra were taken from the ELODIE archive \citep{ELODIEarchive} and continuum was subtracted. We show in Fig. \ref{fig-RTCyg}, hydrogen emission profiles at phase 0.87 which most resemble those in V838 Mon. The most striking similarities are in the close overlap in the span of the H$\gamma$ and H$\delta$ while H$\beta$ occupies the blue half of these short-wavelength lines. In both objects, the profiles show substructures.

\begin{figure}[h!]
    \centering
    \includegraphics[width=1.0\columnwidth,page=1, trim=0 5 0 10]{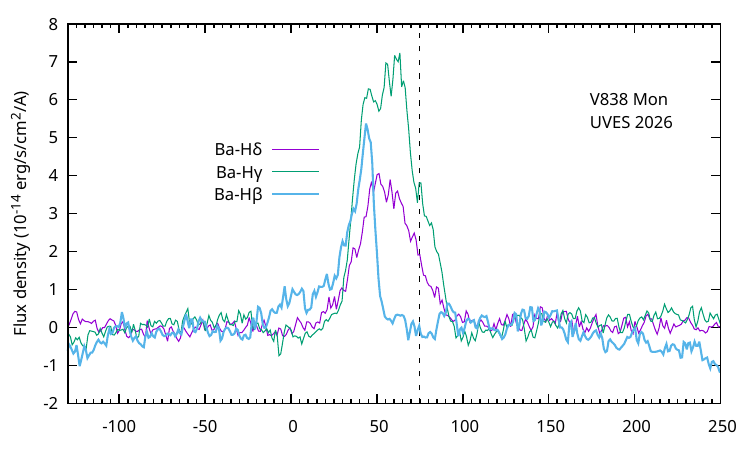}
    \includegraphics[width=1.0\columnwidth,page=1, trim=0 5 0 12]{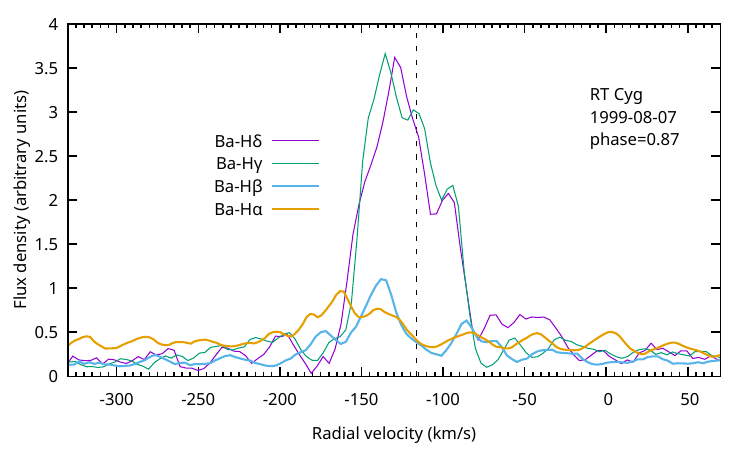}
    \caption{Balmer emission lines in V838 Mon (top) are compared to those in RT Cyg \citep[bottom; from][]{Alvarez2000}. Note that the latter have poorer spectral resolution. Continua were subtracted. The spectrum of RT Cyg was not corrected for ISM extinction. The vertical lines indicate systemic velocities.}
    \label{fig-RTCyg}
\end{figure}

\end{appendix}

\end{document}